\documentclass[10pt,twocolumn]{IEEEtran}
\usepackage[noadjust]{cite}
\usepackage[subsection]{algorithm}
\usepackage{amsmath,amssymb,amsbsy,algorithmic,bm}
\ifCLASSINFOpdf
   \usepackage[pdftex]{graphicx}
   \DeclareGraphicsExtensions{.pdf,.jpeg,.png}
\else
   \usepackage[dvips]{graphicx}
   \DeclareGraphicsExtensions{.eps}
\fi

\newcommand{\Tr}{\mathrm{Tr}}
\newcommand{\vect}{\mathrm{vec}}
\newcommand{\rank}{\mathrm{rank}}
\newcommand{\bs}{\boldsymbol}
\newcommand{\diag}{\mathrm{diag}}
\newtheorem{lemma}{Lemma}
\newtheorem{proposition}{Proposition}

\begin{document}

\title{MIMO Precoding in Underlay Cognitive Radio Systems with Completely Unknown Primary CSI}
\author{Amitav~Mukherjee, Minyan Pei, and A.~Lee~Swindlehurst, \emph{Fellow,~IEEE}
\thanks{The authors are with the Dept.~of EECS,
University of California, Irvine, CA 92697-2625, USA. {\tt (e-mail: \{amukherj; mpei; swindle\}@uci.edu)}}
\thanks{This work was supported by the U.S. ARO under MURI grant W911NF-07-1-0318, and by the NSF under grant CCF-1117983.}
}

\maketitle

\begin{abstract}
This paper studies a novel underlay MIMO cognitive radio (CR) system, where the instantaneous or statistical channel state information (CSI) of the interfering channels to the primary receivers (PRs) is completely unknown to the CR. For the single underlay receiver scenario, we assume a minimum information rate must be guaranteed on the CR main channel whose CSI is known at the CR transmitter. We first show
that low-rank CR interference is preferable for improving the
throughput of the PRs compared with spreading less power over
more transmit dimensions. Based on this observation, we then
propose a rank minimization CR transmission strategy assuming
a minimum information rate must be guaranteed on the CR
main channel. We propose a simple solution referred to as frugal
waterfilling (FWF) that uses the least amount of power required
to achieve the rate constraint with a minimum-rank transmit
covariance matrix. We also present two heuristic approaches that
have been used in prior work to transform rank minimization
problems into convex optimization problems. The proposed schemes are then generalized to an underlay MIMO CR downlink network with multiple receivers. Finally, a theoretical analysis of the interference temperature and leakage rate outage probabilities at the PR is presented for Rayleigh fading channels. We demonstrate
that the direct FWF solution leads to higher PR throughput even
though it has higher interference ``temperature" (IT) compared
with the heuristic methods and classic waterfilling, which calls into question the use of IT
as a metric for CR interference. 
\end{abstract}
\section{INTRODUCTION}

Cognitive radios (CRs) have gained prominence as an efficient method of improving spectrum utilization by allowing coexistence with licensed networks, which is denoted as dynamic spectrum access (DSA). One popular variant of DSA is known as spectrum underlay, where underlay cognitive transmitters (UCTs) operate simultaneously with licensed (primary) users, but adapt their transmission parameters so as to confine the interference perceived at the primary receivers (PRs) to a pre-specified threshold \cite{Jafar09}. Therefore, a fundamental challenge
for the CR is to balance between maximizing its own transmit
rate and minimizing the interference it causes to the PRs.
In CR networks with single-antenna nodes, this is usually
achieved by exploiting some knowledge of the interfering
cross-channels to the PRs at the UCT and performing some
admission control algorithms with power control \cite{Hossain_08}.

 If UCTs are equipped with multiple antennas, the available spatial degrees of freedom can be used to mitigate interference to the PRs during transmission to the underlay receivers. Multi-antenna CR networks have recently received extensive attention, assuming some knowledge of the interfering cross-channels to the PRs at the UCT, either perfect PR cross-channel state information (CSI) \cite{Zhang08}-\cite{Wang10}, perturbed PR CSI \cite{Poor09}-\cite{Mouthaan10}, or statistical PR CSI \cite{Phan09}-\cite{So11}. However, the UCT may not have the luxury of knowing the CSI of the cross links to the PRs, as the primary system would not deliberately coordinate the collection of CSI for the CR system.

 In this work, we consider the novel scenario where both the realizations and distribution of the PR cross-channels are \emph{completely unknown} at the CR, thereby precluding the overwhelming majority of existing spectrum underlay schemes in the literature \cite{Zhang08,Letaief09}. Such a scenario of completely-unknown PR CSI is relevant in a number of instances, for example, when the PR transmits intermittently and therefore stymies attempts to learn the cross-channel, when channels are varying rapidly over time, when the PT and PR do not employ time-division duplexing as assumed in \cite{Yi09,Vorobyov11} among others, or when there are a plurality of active PTs/UCTs and it is impossible to indirectly estimate specific channels.

Specifically, we propose a rank
minimization transmission strategy for the UCT while maintaining a
minimum information rate on the CR link, and we present a simple
solution referred to as frugal waterfilling (FWF) that uses the least
amount of power required to achieve the rate constraint with a
minimum-rank covariance matrix. In the context of MIMO interference
channels (for which the CR underlay network is a special case),
rank-minimization has been shown to be a reinterpretation of
interference alignment \cite{Dimakis10}, but this approach requires
knowledge of interfering cross-channels and treats the overall system
sum rate or degrees-of-freedom as the performance metric, assumptions
which are both markedly different from the underlay CR scenario we
consider.

We also describe two heuristic approaches that have been used in prior
work to transform rank minimization problems (RMP) into problems that
can be solved via convex optimization.  These approaches approximate
the rank objective function with two relaxations, one based on the
nuclear norm \cite{Fazel2001}, and the other on a log-determinant
function \cite{Fazel2003}.  We show theoretically and via numerical
simulation that minimizing the rank of the UCT spatial covariance
matrix leads to the highest PR throughput in general Rayleigh-fading
channels, compared with spreading the transmit power over more
dimensions.  Furthermore, our simulations indicate that FWF provides a
higher PR throughput than the nuclear-norm and log-det heuristic
solutions, even though FWF has a higher interference ``temperature''
(IT).  This suggests that the commonly used IT metric does not
accurately capture the impact of the CR interference on PR
performance.  Instead, we propose a metric based on interference
leakage (IL) rate that more accurately reflects the influence of the
CR interference.


 This paper is organized as follows. The underlay system model is introduced in Section~\ref{sec:SysModel}. PR CSI-unaware UCT transmit strategies for a single UCR are presented in Section~\ref{sec:BlindPrec}. The generalization to the underlay downlink with multiple UCRs is shown in Section~\ref{sec:CRdownlink}. A random matrix-theoretic analysis of the primary outage probability due to the proposed strategies is given in Section~\ref{sec:PR_OutP}. The penultimate Section~\ref{sec:sim} presents numerical simulations for various underlay scenarios, and we conclude in Section~\ref{sec:concl}.

 \emph{Notation}:
We will use $\mathcal{CN}(\mathbf{0},\mathbf{Z})$ to denote a circularly symmetric complex
Gaussian distribution with zero mean and covariance matrix $\mathbf{Z}$,
$\mathcal{E}\{\cdot\}$ to denote expectation, $\vect(\cdot)$ the matrix column stacking operator, $(\cdot)^T$ the transpose, $(\cdot)^H$ the Hermitian
transpose, $(\cdot)^{-1}$ the matrix inverse, $\Tr(\cdot)$ the
trace operator, $\left| \cdot \right|$ or $\det$ the matrix determinant, $\diag(\mathbf{a})$ a diagonal matrix with the elements of $\mathbf{a}$ on the main diagonal, $|\mathbf{A}|_{i,j}$ the $(i,j)$ element of $\mathbf{A}$, $\Gamma(x)$ the gamma function, and $\mathbf{I}$ is the identity matrix.

\section{System Model}\label{sec:SysModel}
\begin{figure}[htbp]
\centering
\includegraphics[width=\linewidth]{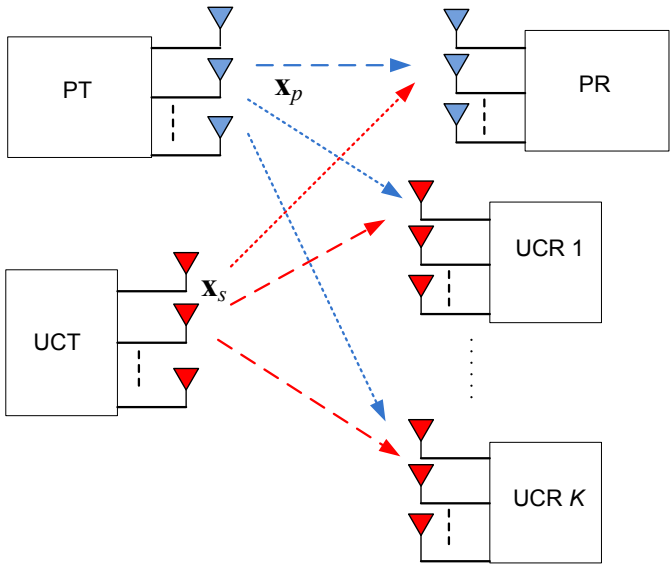}
\caption{Cognitive radio network with a multi-antenna underlay CR transmitter, $K$ underlay receivers, and a single MIMO PT-PR pair.}
\label{fig_network}
\end{figure}
A generic MIMO underlay CR network with $K$ multi-antenna underlay receivers is shown in Fig.~1, where a primary system and an underlay CR system share the same spectral band. Since the UCT transmit strategies are independent of the cross-channels to primary users, the numbers of PRs and PTs and their array sizes can be made arbitrary; however to simplify notation we will consider a solitary multi-antenna PT-PR pair. To introduce the problem we first consider the scenario with a single UCR, and generalize to the case of $K>1$ in Sec.~\ref{sec:CRdownlink}.

We consider a multi-antenna UCT equipped with $N_a$ antennas, which transmits a signal vector $\mathbf{x}_s \in \mathbb{C}^{N_a \times 1}$, to its $N_s$-antenna underlay cognitive receiver (UCR).
 The PT is equipped with $N_p$ antennas and transmits signal $\mathbf{x}_{p}\in \mathbb{C}^{N_p \times 1}$ to the $N_r$-antenna PR.
Thus, the UCR observes
\begin{equation}
{{\mathbf{y}}_s} = {{\mathbf{G}}_1}{{\mathbf{x}}_s} + {{\mathbf{G}}_2}{{\mathbf{x}}_p} + {{\mathbf{n}}_s},
\end{equation}
where $\mathbf{G}_1 \in {\mathbb{C}^{{N_s} \times {N_a}}},\mathbf{G}_2\in {\mathbb{C}^{{N_s} \times {N_p}}}$ are the complex MIMO channels from the UCT and PT, 
and $\mathbf{n}_{s}\sim \mathcal{CN}(\mathbf{0},\sigma_s^2\mathbf{I})$ is complex additive white Gaussian noise. We assume Gaussian signaling with zero mean and second-order statistics $\mathcal{E}\{\mathbf{x}_s\mathbf{x}_s^H\} = \mathbf{Q}_s$, and the average UCT transmit power is assumed to be bounded:
\begin{equation*}
\Tr(\mathbf{Q}_s) \le P_s.
\end{equation*}
The signal at the PR is given by
\begin{equation}
\mathbf{y}_{p}=\mathbf{H}_1{\mathbf{x}}_{p}+\mathbf{H}_2\mathbf{x}_s+\mathbf{n}_{p},
\end{equation}
where $\mathbf{H}_1\in {\mathbb{C}^{{N_p} \times {N_r}}},\mathbf{H}_2\in {\mathbb{C}^{{N_r} \times {N_a}}}$ are the channels from the PU and CR transmitters (assumed to be full-rank), and $\mathbf{n}_{p}\sim \mathcal{CN}(\mathbf{0},\sigma_p^2\mathbf{I})$ is complex additive white Gaussian noise. The primary signal is also modeled as a zero-mean complex Gaussian signal with covariance matrix ${{{\mathbf{Q}}_p}}$ and average power constraint $\Tr(\mathbf{Q}_p) \le P_p$. We will assume ${{{\mathbf{Q}}_p}}$ is fixed and the channels are mutually independent and each composed of i.i.d. zero-mean circularly symmetric complex Gaussian entries, and focus our attention on the design of the UCT transmit signal.

We assume there is no cooperation between the PT and UCT during transmission, and that both receivers treat interfering signals as noise. The network is essentially an asymmetric 2-user MIMO interference channel, where the UCT attempts to minimize the interference to the PR, but no such reciprocal gesture is made by the PT.
The interference covariance matrix at the PR is
\begin{equation}
\mathbf{K}_p = \mathbf{H}_{2}\mathbf{Q}_s\mathbf{H}_{2}^H.
\end{equation}
Define the interference temperature at the PR as \cite{Zhang08}-\cite{So11}
\begin{equation}\label{eq:IntTemperature}
{T_p}\left( {\mathbf{Q}_s} \right)=\Tr\left(\mathbf{K}_p\right).
\end{equation}
Without knowledge of $\mathbf{H}_2$ or its distribution, the UCT cannot directly optimize the PR interference temperature or outage probability as in existing underlay proposals \cite{Zhang08}-\cite{So11}. To our best knowledge, precoding strategies and performance analyses for MIMO underlay systems with completely unknown primary CSI have not been presented in the literature thus far. In addition to \cite{Zhang08}-\cite{Noam12} not being applicable, the blind interference alignment method for the 2-user MIMO interference channel \cite{Jafar12} is also precluded since it requires knowledge of the cross-channel coherence intervals, which we assume is also unknown.  In \cite{Noam12}, a blind underlay precoding scheme is proposed where the MIMO CR iteratively updates its spatial covariance by observing the transmit power of a solitary PT. The UCT attempts to infer the least-harmful spatial orientation towards the PR, but requires that the PT employ a power control scheme monotonic in the interference caused by the CR, and that the cross-channel remains constant during the learning process. In contrast, we investigate simple non-iterative CR precoding strategies which do not impose any restrictions on the PT transmission strategy or number of PTs, or cross-channel coherence intervals.


The PT achieves the following rate on its link:
\begin{equation}\label{eq:PURate}
{R_p}\left( {{{\mathbf{Q}}_s}} \right) = {\log _2}\left| {{\mathbf{I}} + {{\mathbf{H}}_1}{{\mathbf{Q}}_p}{\mathbf{H}}_1^H{{{\left( {{{\mathbf{K}}_p} + \sigma _p^2{\mathbf{I}}} \right)}^{ - 1}}}} \right|.
\end{equation}
Similarly, the achievable rate on the CR link is
\begin{equation}\label{eq:CRRate}
{R_s}\left( {{{\mathbf{Q}}_s}} \right) = {\log _2}\left| {{\mathbf{I}} + {{\mathbf{G}}_1}{{\mathbf{Q}}_s}{\mathbf{G}}_1^H{{{\left( {{{\mathbf{K}}_s} + \sigma _s^2{\mathbf{I}}} \right)}^{ - 1}}}} \right|
\end{equation}
where ${{\mathbf{K}}_s} = {{\mathbf{G}}_2}{{\mathbf{Q}}_p}{\mathbf{G}}_2^H$ represents the interference from the PT.

\section{A Rank Minimization strategy for CSI-Unaware Underlay Transmission}\label{sec:BlindPrec}

We now expound on the fundamental motivation underlying the UCT transmission strategies proposed in this work. As we have seen, due to a lack of knowledge of $\mathbf{H}_2$ or its distribution, the UCT cannot directly optimize the PU interference temperature. Hence, we propose an alternative transmission strategy where the UCT tries to minimize a measure of the interference caused to the PR in a ``best-effort" sense, while achieving a target data rate to the UCR. Assuming that $\mathbf{Q}_p$ is fixed, we first show that in the clairvoyant case where the UCT has some knowledge of the channel to the PR ($\mathbf{H}_2$), a rank-1 UCT covariance matrix $\mathbf{Q}_s$ causes least interference to the primary link, which is described in the following proposition.

\begin{proposition}
The optimal solution to the clairvoyant problem
\begin{subequations}
\begin{align}
\max\limits_{\mathbf{Q}_s} &\quad \mathcal{E}\left\{R_p(\mathbf{Q}_s)\right\}  \\
\mathrm{s.t.} & \quad {R_s}\left( {{{\mathbf{Q}}_s}} \right) = {R_b} \\
& \quad \Tr \left( {{{\mathbf{Q}}_s}} \right) \leq {P_s}\\
& \quad \mathbf{Q}_s \succeq \mathbf{0}.
\end{align}
\end{subequations}
that maximizes the average PR rate is of rank one, i.e., $\rank(\mathbf{Q}_s^{\star})=1$.
\end{proposition}
\begin{IEEEproof}
Since $\mathbf{Q}_s \succeq \mathbf{0}$, it can be expressed as $\mathbf{Q}_s = \mathbf{U}\bs{\Lambda}\mathbf{U}^H$, where $\bs{\Lambda}$ is the diagonal matrix of eigenvalues of $\mathbf{Q}_s$ and $\mathbf{U}$ is the unitary matrix with columns consisting of the eigenvectors of $\mathbf{Q}_s$. Defining $\tilde{\mathbf{H}}_2 = \mathbf{H}_2\mathbf{U}$, it follow from Lemma 5 in \cite{Telatar1999} that the distribution of $\tilde{\mathbf{H}}_2$ is the same as that of $\mathbf{H}_2$. As a result, the average PU rate can be expressed as
\begin{equation*}
\begin{split}
&\quad R_p(\mathbf{Q}_s) = \Phi(\bs{\Lambda})\\
& = \mathcal{E} \bigg\{ \log_2 \bigg[\det \left(\mathbf{I} + \mathbf{H}_1\mathbf{Q}_p \mathbf{H}_1^H {\left( \tilde{\mathbf{H}}_2\bs{\Lambda}\tilde{\mathbf{H}}_2^H + \sigma_p^2 \mathbf{I}\right)}^{ - 1} \right)\bigg] \bigg\}
\end{split}
\end{equation*}
Thus, the problem we considered is essentially equivalent to constructing the diagonal matrix $\bs{\Lambda}$ with real nonnegative entries so as to maximize $\Phi(\bs{\Lambda})$ under the constraint $\Tr(\bs{\Lambda})= P_s$.

From \cite[Lemma 3]{Kashyap2004},\cite{Mao06}, we have that $\Phi(\bs{\Lambda})$ is a convex function of $\bs{\Lambda}$. Note that given any permutation matrix $\bs{\Pi}$, we see (using Lemma 5 in \cite{Telatar1999} again) that
\begin{equation*}
\Phi (\bs{\Pi\Lambda}\bs{\Pi}^{H} ) = \Phi(\bs{\Lambda}).
\end{equation*}
From convexity, we have
\begin{equation*}
\Phi\left(\frac{1}{N_a!}\sum_{\bs{\Pi}}\bs{\Pi}\bs{\Lambda}\bs{\Pi}^{H}\right) \leq \frac{1}{N_a!}\sum_{\bs{\Pi}} \Phi(\bs{\Pi}\bs{\Lambda}\bs{\Pi}^{H}) = \Phi(\bs{\Lambda})
\end{equation*}
where we have used Jensen's inequality. From the transmit power constraint, we have $\frac{1}{N_a!}\sum_{\bs{\Pi}}\bs{\Pi}\bs{\Lambda}\bs{\Pi}^{H} = (P_s/N_a)\mathbf{I}_{N_a}$. Thus, we have proved that the least PU rate is obtained by $\bs{\Lambda}=(P_s/N_a)\mathbf{I}_{N_a}$. Further, due to convexity, we can argue that the largest PU rate is obtain by a point farthest away from $\bs{\Lambda}=(P_s/N_a)\mathbf{I}_{N_a}$. Thus, we want $\bs{\Lambda}^{\star}=\diag(\lambda^{\star}_1,\dots,\lambda^{\star}_{N_a})$ that satisfy \cite{Blum2003}
\begin{equation*}
\max \sum_{i=1}^{N_a} \left(\lambda_i-\frac{P_s}{N_a} \right)^2 ~~~~ \mathrm{s.~t.~}  \sum_{i=1}^{N_a}\lambda_i = P_s
\end{equation*}
Now,
\begin{equation*}
\begin{split}
\sum_{i=1}^{N_a} \left(\lambda_i-\frac{P_s}{N_a} \right)^2 & = \sum_{i=1}^{N_a} \lambda_i^2 - 2\frac{P_s}{N_a}\sum_{i=1}^{N_a} \lambda_i + \frac{P_s^2}{N_a} \\
& = P_s^2\left( \sum_{i=1}^{N_a} \left(\frac{\lambda_i}{P_s}\right)^2- \frac{1}{N_a} \right)\\
& \leq P_s^2\left( \sum_{i=1}^{N_a} {\frac{\lambda_i}{P_s}} - \frac{1}{N_a} \right) \\
& = P_s^2 \left(1- \frac{1}{N_a} \right)
\end{split}
\end{equation*}
where we used $\sum_{i=1}^{N_a} \left(\frac{\lambda_i}{P_s}\right)^2 \leq \sum_{i=1}^{N_a} {\frac{\lambda_i}{P_s}} =1$, and the equality is satisfied by any $(\lambda^{\star}_1,\dots,\lambda^{\star}_{N_a})$ with all zeros except for one nonzero entry.
Hence, we conclude that $\rank(\mathbf{Q}_s^{\star})=\rank(\bs{\Lambda}^{\star})=1$.
\end{IEEEproof}

Therefore, in the clairvoyant case where the UCT has some knowledge of the primary CSI, a rank-1 $\mathbf{Q}_s$ causes least interference to the primary link and full-rank $\mathbf{Q}_s$ causes most interference\footnote{This notion has been echoed in prior art on MIMO interference channels \cite{Andrews07,Arslan07}}. Of course, the optimal $\mathbf{Q}_s$ will depend on the PR CSI, which we have assumed is
unavailable. Still, the result motivates the use of a low-rank
transmit covariance at the UCT. It is evident that the UCT does not actually require knowledge of the PR CSI to minimize the rank of $\mathbf{Q}_s$ needed to achieve a rate target $R_b$ on the CR link. To exploit this observation, we henceforth pose the UCT precoder design problem when the PR CSI is completely unknown as
\begin{subequations}\label{pr:RMP}
\begin{align}
(\textrm{P0}):\quad \min &\quad \rank  \left( {{{\mathbf{Q}}_s}} \right)  \\
\mathrm{s.t.} & \quad {R_s}\left( {{{\mathbf{Q}}_s}} \right) = {R_b} \\
& \quad \Tr \left( {{{\mathbf{Q}}_s}} \right) \leq {P_s}\\
& \quad \mathbf{Q}_s \succeq \mathbf{0}.
\end{align}
\end{subequations}
This is a rank-minimization problem (RMP), and in general is computationally hard to solve since the $\rank$ function is quasi-concave and not convex. As
explained below, however, in this case a simple waterfilling
solution can be obtained. In \cite{Yu2011}, the mutual information of the CR link is maximized subject to an interference temperature constraint and arbitrary transmit covariance rank constraints, which implies knowledge of PR CSI and thus differs from this work.

\section{Solutions for the Rank Minimization Design}\label{sec:Heuristics}
Notice that the design problem (P0) is ill-posed in the sense that there are potentially an infinite number of solutions. Suppose that we find one minimum-rank solution to (P0) such that $\mathbf{Q}_s=\mathbf{U}\mathbf{\Lambda}\mathbf{U}^{H}$ for some unitary matrix $\mathbf{U}$ and diagonal matrix $\mathbf{\Lambda}$ satisfies $R_s(\mathbf{Q}_s)=R_b$, $\Tr(\mathbf{\Lambda})\leq P_s$. If the required power $\Tr(\mathbf{\Lambda})$ is strictly smaller than $P_s$ , then we could find an infinite number of solutions by making small perturbations to $\mathbf{U}$, which while leading to a higher power requirement, still would require less power than $P_s$.  Obviously, the solution with least power is more desirable for our underlay CR system in order to minimize the interference caused to the PR, and this solution can easily be found using the \emph{Frugal Waterfilling} (FWF) approach described next.

\subsection{FWF Approach}\label{sec:FWF}

The FWF solution seeks to find the least amount of power required to
achieve the CR rate target of $R_b$ with the minimum rank transmit
covariance $\mathbf{Q}_s$.  The optimization problem can be solved
using a combination of the classic waterfilling (CWF) algorithm and a
simple bisection line search. The description for FWF is outlined as
Algorithm~IV-A.1 below.  In brief, FWF cycles through the possible
number of transmit dimensions in ascending order starting with a
rank-one $\mathbf{Q}_s$, and at each step computes the transmit power
required to meet the rate constraint $R_b$ based on CWF.  This requires
a simple line search over the transmit power for each step.  Once
a solution is found that satisfies the transmit power constraint,
the algorithm terminates.  If no feasible solution is found for
all $N_a$ transmit dimensions, the CR link will be in outage.
\begin{algorithm}[htpb]
\caption{Frugal Waterfilling for UCT Rank/Power Tradeoff \cite{Mukherjee12}}
\label{alg1}
\begin{algorithmic}
\REQUIRE $P_s > 0, R_b > 0$
\STATE set $r=\rank(\mathbf{G}_1)$
\FOR{$M=1$ to $r$}
\STATE Solve:
\begin{equation*}
\begin{array}{c}
p(M) = \min \Tr(\mathbf{Q}_s) \\[10pt]
\mbox{\rm s.~t.} \; \;  {\log _2}\left| {{\mathbf{I}} + {{\mathbf{G}}_1}{{\mathbf{Q}}_s}{\mathbf{G}}_1^H{{{\left( {{{\mathbf{K}}_s} + \sigma _s^2{\mathbf{I}}} \right)}^{ - 1}}}} \right|=R_b \; .
\end{array}
\end{equation*}
\ENDFOR
\IF{$p(r) > P_s$}
\STATE Declare outage
\ELSE
\STATE {\underline{CWF solution}}: ${\displaystyle N = arg\min_M p(M)};$
\STATE {\underline{FWF solution}}: ${\displaystyle N = arg\min_M M};$
\STATE $\mathbf{Q}_s$ determined by waterfilling $p(N)$ over $N$ largest singular
values of ${{\left( {{{\mathbf{K}}_s} + \sigma _s^2{\mathbf{I}}} \right)}^{ -1/2}}\mathbf{G}_1$
\ENDIF
\end{algorithmic}
\end{algorithm}

The FWF algorithm was presented in brief without analysis by the authors in \cite{Mukherjee12}, and through simulation were shown to be an effective transmission strategy in conventional downlink, wiretap, and underlay networks.
While FWF finds an efficient solution to (P0), in general rank
minimization problems are difficult to solve and often require
exponential-time complexity.  Consequently, heuristic approximations
to the matrix rank have been proposed as alternatives in order to
yield simpler optimization problems.  In particular, the nuclear
norm \cite{Fazel2001} and log-determinant \cite{Fazel2003} heuristics
have been proposed in order to convexify RMP problems like (P0)
and provide approximate solutions with polynomial-time complexity.
In the discussion below, we show how these approximations can be
applied to the RMP we consider in this paper.

\subsection{Nuclear Norm and Log-det Heuristic}\label{sec:Heuristic}
The \emph{nuclear norm heuristic} is based on the fact the nuclear
norm (sum of the singular values of a matrix) is the convex envelope
of the rank function on the unit ball.  When the matrix is positive
semidefinite, the nuclear norm is the same as the trace function. As a
result, the design problem (P0) can be formulated as follows:
\begin{equation}\label{pr:NuclearNormHeuristic}
\begin{split}
(\textrm{P1}): \quad \min &\quad \Tr \left( \mathbf{Q}_s \right)  \\
\mathrm{s.t.} & \quad {R_s}\left( {{{\mathbf{Q}}_s}} \right) = {R_b} \\
& \quad \Tr \left( {{{\mathbf{Q}}_s}} \right) \leq {P_s}\\
& \quad \mathbf{Q}_s \succeq \mathbf{0} \; .
\end{split}
\end{equation}
The nuclear norm heuristic (P1) is a convex optimization problem and
can be solved using the CWF algorithm together with a bisection line
search (similar to FWF). It is well known that under the CWF algorithm, the
lowest transmit power is achieved when $\rank(\mathbf{Q}_s)$ is chosen
as large as possible (up to $\rank(\mathbf{G}_1)$). This is clearly
contrary to the rank-minimization design formulation, which indicates
that the nuclear norm approach for this problem is a poor approximation.

Using the function $\log\det(\mathbf{Q}_s+\delta\mathbf{I})$ as a smooth surrogate for $\rank(\mathbf{Q}_s)$, the \emph{log-det heurisitic} can be described as follows:
\begin{equation}\label{pr:Log-detHeuristic}
\begin{split}
(\textrm{P2}):\quad \min &\quad \log\det \left( \mathbf{Q}_s + \delta\mathbf{I} \right)  \\
\mathrm{s.t.} & \quad {R_s}\left( {{{\mathbf{Q}}_s}} \right) = {R_b} \\
& \quad \Tr \left( {{{\mathbf{Q}}_s}} \right) \leq {P_s}\\
& \quad \mathbf{Q}_s \succeq \mathbf{0},
\end{split}
\end{equation}
where $\delta \ge 0$ can be interpreted as a small regularization
constant (we choose $\delta = 10^{-6}$ for numerical examples).  Since
the surrogate function $\log\det(\mathbf{Q}_s+\delta\mathbf{I})$ is
smooth on the positive definite cone, it can be minimized
using a local minimization method. We use iterative
linearization to find a local minimum to the optimization problem (P2)
\cite{Fazel2003}. Let $\mathbf{Q}_s^{(k)}$ denote the $k$th iteration
of the optimization variable $\mathbf{Q}_s$. The first-order Taylor series expansion of $\log\det(\mathbf{Q_s}+\delta\mathbf{I})$ about $\mathbf{Q}_s^{(k)}$ is given by
\begin{equation}\label{eq:linear}
\begin{split}
\log\det(\mathbf{Q}_s+\delta\mathbf{I})~\approx~ & \log\det(\mathbf{Q}_s^{(k)}+\delta\mathbf{I})+\\
& \Tr\big[(\mathbf{Q}_s^{(k)}+\delta\mathbf{I})^{-1}(\mathbf{Q}_s-\mathbf{Q}_s^{(k)})\big].
\end{split}
\end{equation}
Hence, we could attempt to minimize $\log\det(\mathbf{Q}_s+\delta\mathbf{I})$ by iteratively minimizing the local linearization (\ref{eq:linear}). This leads to
\begin{equation}\label{eq:iteration}
\mathbf{Q}_s^{(k+1)} = \mathrm{argmin} ~ \Tr \big[(\mathbf{Q}_s^{(k)}+\delta\mathbf{I})^{-1}\mathbf{Q}_s \big].
\end{equation}
If we choose $\mathbf{Q}_s^{(0)}=\mathbf{I}$, the first iteration of (\ref{eq:iteration}) is equivalent to minimizing the trace of $\mathbf{Q}_s$. Therefore, this heuristic can be viewed as a refinement of the nuclear norm heuristic. As a result, we always pick $\mathbf{Q}^{(0)}_s=\mathbf{I}$, so that $\mathbf{Q}^{(1)}_s$ is the result of the trace heuristic, and the iterations that follow try to reduce the rank of $\mathbf{Q}^{(1)}_s$ further.

Note that at each iteration we will solve a weighted trace minimization problem, which is equivalent to the following optimization problem
\begin{equation}\label{pr:Log-detIter}
\begin{split}
(\textrm{P2-1}): \quad \min &\quad \Tr \left( \mathbf{F}^H \mathbf{A}\mathbf{F} \right)  \\
\mathrm{s.t.} & \quad \log_2\det \left(\mathbf{I}+\mathbf{F}^H \mathbf{R}\mathbf{F} \right) = {R_b} \\
& \quad \Tr \left( \mathbf{F}^H \mathbf{F} \right) \leq {P_s}.
\end{split}
\end{equation}
where $\mathbf{A}=(\mathbf{Q}_s^{(k)}+\delta\mathbf{I})^{-1}$, $\mathbf{R}=\mathbf{G}_1^{H}(\mathbf{G}_2\mathbf{Q}_p\mathbf{G}_2^{H}+\sigma_s^2\mathbf{I})^{-1}\mathbf{G}_1$.
This is a Schur-concave optimization problem with multiple trace/log-det constraints. From Theorem 1 in \cite{Yu2011}, the optimal solution to problem \eqref{pr:Log-detIter} is $\mathbf{F}^{\star}=\mathbf{A}^{-1/2}\mathbf{U}\mathbf{\Sigma}$, and $\mathbf{Q}_s^{(k+1)}=\mathbf{F}^{\star}{\mathbf{F}^{\star}}^{H}$ is an optimal solution to (\ref{eq:iteration}), where $\mathbf{A}^{-1/2}=\mathbf{U}_{\mathbf{A}}\mathbf{\Lambda}^{-1/2}_{\mathbf{A}}\mathbf{U}^{H}_{\mathbf{A}}$, $\mathbf{U}_{\mathbf{A}}$ and $\mathbf{\Lambda}_{\mathbf{A}}$ are defined in the eigen-decomposition $\mathbf{A}=\mathbf{U}_{\mathbf{A}}\mathbf{\Lambda}_{\mathbf{A}}\mathbf{U}^{H}_{\mathbf{A}}$, $\mathbf{U}$ is a unitary matrix, and $\mathbf{\Sigma}=\diag(\sqrt \mathbf{p})$ is a rectangular diagonal matrix.

Substituting the optimal solution structure $\mathbf{F}^{\star}$ into \eqref{pr:Log-detIter}, we have the following equivalent problem
\begin{equation}\label{pr:Log-detEqu}
\begin{split}
(\textrm{P2-2}): \quad \min &\quad \Tr \left( \mathbf{\Sigma}\mathbf{\Sigma}^{H} \right)  \\
\mathrm{s.t.} & \quad \log_2\det \left(\mathbf{I}+\mathbf{\Sigma}^H \mathbf{U}^{H}\tilde{\mathbf{R}}\mathbf{U}\mathbf{\Sigma} \right) = {R_b} \\
& \quad \Tr \left( \mathbf{U}^H \mathbf{A}^{-1} \mathbf{U} \mathbf{\Sigma}\mathbf{\Sigma}^{H} \right) \leq {P_s}
\end{split}
\end{equation}
where $\tilde{\mathbf{R}}=\mathbf{A}^{-1/2}\mathbf{R}\mathbf{A}^{-1/2}$. It is found that the equivalent problem \eqref{pr:Log-detEqu} is essentially equivalent to the converse formulation
\begin{equation}\label{pr:Log-detEqu1}
\begin{split}
(\textrm{P2-3}): \quad \max &\quad \log_2\det \left(\mathbf{I}+\mathbf{\Sigma}^H \mathbf{U}^{H}\tilde{\mathbf{R}}\mathbf{U}\mathbf{\Sigma} \right) \\
\mathrm{s.t.} & \quad \Tr \left( \mathbf{\Sigma}\mathbf{\Sigma}^{H} \right) = P_0 \\
& \quad \Tr \left( \mathbf{U}^H \mathbf{A}^{-1} \mathbf{U} \mathbf{\Sigma}\mathbf{\Sigma}^{H} \right) \leq {P_s}.
\end{split}
\end{equation}
This is because both formulations (\ref{pr:Log-detEqu}) and (\ref{pr:Log-detEqu1}) describe the same tradeoff curve of performance versus power. Therefore, the quality-constrained problem (P2-2) can be numerically solved by iteratively solving the power-constrained problem (P2-3), combined with the bisection method.

The problem formulation in (P2-3) is a Schur-convex optimization problem with two trace constraints. Using Theorem 1 in \cite{Yu2011} again, if we let $\tilde{\mathbf{R}}=\mathbf{U}_{\tilde{\mathbf{R}}}\mathbf{\Lambda}_{\tilde{\mathbf{R}}}\mathbf{U}_{\tilde{\mathbf{R}}}^{H}$ denote the eigen-decomposition of $\tilde{\mathbf{R}}$, then the optimal unitary matrix $\mathbf{U}$ will be chosen as $\mathbf{U}_{\tilde{\mathbf{R}}}$. Denoting $\mathbf{a}=\diag(\mathbf{U}^{H}\mathbf{A}^{-1}\mathbf{U})$ and letting $\lambda_{\tilde{\mathbf{R}},1} \geq \lambda_{\tilde{\mathbf{R}},2} \geq \cdots \geq \lambda_{\tilde{\mathbf{R}},N_a}$ represent the diagonal elements of $\mathbf{\Lambda}_{\tilde{\mathbf{R}}}$, the optimal power allocation can be shown to have the form of a \emph{multilevel} waterfilling solution:
\begin{equation}
p_i = \left(\frac{1}{\mu + a_i\nu} - \frac{1}{\lambda_i} \right)^{+}, ~~ i = 1,\dots,N_a
\end{equation}
where $a_i$ is the $i$th element of $\mathbf{a}$, and $\mu, \nu$ can be shown to be the nonnegative Lagrange multipliers associated with the two power constraints. The algorithmic description for the log-det heuristic approach is outlined in Algorithm~\ref{alg2}.

\begin{algorithm}[htpb]
\caption{Iterative log-det heuristic Algorithm for rank-minimization problem} \label{alg2}
\begin{algorithmic}
\REQUIRE $P_s > 0, R_b > 0$,
  \STATE set $\delta = 10^{-6}, \Delta = 10^{-3}, k=0$,
  \STATE ~~~ $\mathbf{Q}_s^{(0)}=\mathbf{I}$, $\mathbf{R}=\mathbf{G}_1^{H}(\mathbf{K}_s+\sigma_s^2\mathbf{I})^{-1}\mathbf{G}_1$.
\REPEAT
  \STATE $\mathbf{A}=(\mathbf{Q}_s^{(k)}+\delta\mathbf{I})^{-1}$, $\tilde{\mathbf{R}}=\mathbf{A}^{-1/2}\mathbf{R}\mathbf{A}^{-1/2}$,
  \STATE $\operatorname{eig}(\tilde{\mathbf{R}})=\mathbf{U}_{\tilde{\mathbf{R}}}\mathbf{\Lambda}_{\tilde{\mathbf{R}}}\mathbf{U}_{\tilde{\mathbf{R}}}^{H}$. \STATE set $\mathbf{U}=\mathbf{U}_{\tilde{\mathbf{R}}}$, $\mathbf{a}=\diag(\mathbf{U}^{H}\mathbf{A}^{-1}\mathbf{U})$.
\STATE Solve:
  \begin{equation*}
  \begin{split}
  \min ~~ & \mathbf{1}^{T}\mathbf{p} \\
  \mbox{\rm s.~t.} ~~ & \log_2 \left(\prod_{i}^{} (1+p_i \lambda_{\tilde{\mathbf{R}},i})\right)=R_b \\
  & \mathbf{a}^T\mathbf{p} \leq {P_s}.
  \end{split}
  \end{equation*}
\STATE $\mathbf{\Sigma}=\diag(\mathbf{p})$
\STATE $\mathbf{F}=\mathbf{A}^{-1/2}\mathbf{U}\mathbf{\Sigma}$
\STATE $\mathbf{Q}_s^{(k+1)}=\mathbf{F}\mathbf{F}^{H}$
\UNTIL {$\log_2(\det(\mathbf{Q}_s^{(k)}+\delta \mathbf{I}))-\log_2(\det(\mathbf{Q}_s^{(k+1)}+\delta \mathbf{I})) < \Delta$}
\IF{$\mathbf{a}^T \mathbf{p} > P_s$}
\STATE Declare outage
\ELSE
\STATE $\rho = \mathbf{a}^T \mathbf{p} /P_s;$ $\quad \mathbf{Q}_s = \mathbf{Q}_s^{(k+1)}$
\ENDIF
\end{algorithmic}
\end{algorithm}

\section{Underlay CR Downlink}\label{sec:CRdownlink}
In this section we extend the blind underlay precoding paradigm to a MIMO underlay downlink network with $K$ UCRs. We consider a modified block-diagonalization precoding strategy \cite{SwindleBD} where multiple data streams are transmitted to each UCR. Let each UCR be equipped with $N_s$ antennas for simplicity, although the proposed precoding schemes hold for heterogeneous receiver array sizes as long as the total number of receive antennas does not exceed $N_a$. The extension to the case where the UCT serves $N_a$ spatial streams regardless of the total number of receive antennas can be made using the coordinated beamforming approach \cite{SwindleBD}, for example. The received signal at UCR $k$ is now
\begin{equation}
{{\mathbf{y}}_k} = {{\mathbf{G}}_{k,1}}{{\mathbf{W}}_k}{{\mathbf{s}}_{u,k}} + \sum\limits_{j \ne k}^{{K_u}} {{{\mathbf{G}}_{k,1}}{{\mathbf{W}}_j}{{\mathbf{s}}_{u,j}}}  + {{{\mathbf{G}}_{k,2}}{{\mathbf{s}}_{p}}}  + {{\mathbf{n}}_k}
\end{equation}
where $\mathbf{G}_{k,1} \in \mathbb{C}^{N_s \times N_a}$ is the main channel, ${{\mathbf{W}}_k}\in \mathbb{C}^{N_a \times l_k}$ is the precoding matrix applied to signal $\mathbf{s}_{u,k} \in \mathbb{C}^{l_k \times 1}$ for user $k$, $\mathbf{s}_{p}$ is the PT signal received over interfering channel ${\mathbf{G}}_{k,2}\in \mathbb{C}^{N_s \times N_p}$, and ${{\mathbf{n}}_k}\sim \mathcal{CN}(0,\sigma_k^2 \mathbf{I})$ is additive Gaussian noise. The UCT transmit covariance per UCR is now $\mathbf{Q}_{k,s}={\mathbf{W}}_k{\mathbf{W}}_k^H$, and the overall UCT transmit covariance assuming independent messages is $\mathbf{Q}_{s}=\sum\nolimits_{k=1}^K \mathbf{Q}_{k,s}$.

We assume each UCR has a desired information rate of $R_k$, and adopt the ``BD for power control" approach in \cite[Sec.~II-B]{SwindleBD}.
Letting ${{\mathbf{W}}_k} = {{\mathbf{T}}_k}{\mathbf{\Lambda }}_k^{{1 \mathord{\left/
 {\vphantom {1 2}} \right.
 \kern-\nulldelimiterspace} 2}}$, it is possible to separately design the beamforming matrix ${\mathbf{T}}_k$ and diagonal power allocation matrix ${\mathbf{\Lambda }}_k$ per user to achieve rate $R_k$ in a two-step process. Let
\[{{\mathbf{G}}_{ - k}} = \left[ {\begin{array}{*{20}{l}}
  {{{\mathbf{G}}_{1,1}}}& \cdots &{{{\mathbf{G}}_{k - 1,1}}}&{{{\mathbf{G}}_{k + 1,1}}}& \cdots &{{{\mathbf{G}}_{{K,1}}}}
\end{array}} \right]
\]
 represent the the overall UCR downlink channel excluding the $k^{th}$ user. First, a closed-form solution for the unit-power beamforming matrix ${\mathbf{T}}_k$ of user $k$ is obtained from the nullspace of ${{\mathbf{G}}_{ - k}}$. To achieve this, from the SVD ${{\mathbf{G}}_{ - k}} = {{\mathbf{U}}_{ - k}}{{\mathbf{\Sigma }}_{ - k}}{\left[ {\begin{array}{*{20}{c}}
  {{{\mathbf{V}}_{ - k,1}}}&{{{\mathbf{V}}_{ - k,0}}}
\end{array}} \right]^H}$, the last $(N_a-l_k)$ right singular vectors contained in ${\mathbf{V}}_{ - k,0}$ can be used to construct $\mathbf{T}_k$ \cite{SwindleBD}. The BD strategy therefore completely eliminates intra-UCR interference on the underlay downlink, and the residual interference-plus-noise covariance matrix at UCR $k$ is
\begin{equation}
{{\mathbf{Z}}_k} = {{\mathbf{G}}_{k,2}}{{\mathbf{Q}}_p}{\mathbf{G}}_{k,2}^H + \sigma _s^2{\mathbf{I}}.
\end{equation}

 Proceeding to the power allocation step, let $\rank({{\mathbf{G}}_{k,1}}{{\mathbf{T}}_k})=r_k$ for user $k$'s effective channel, and assume $l_k=r_k$. Consider the SVD of user $k$'s pre-whitened effective channel
 \[
 {\mathbf{Z}}_k^{ - {1 \mathord{\left/
 {\vphantom {1 2}} \right.
 \kern-\nulldelimiterspace} 2}}{{\mathbf{G}}_{k,1}}{{\mathbf{T}}_k} = {{\mathbf{U}}_k}{{\mathbf{\Lambda }}_k}{\mathbf{V}}_k^H\]
 where 
 $\mathbf{\Lambda}_k=\operatorname{diag}\left(\lambda_{k,1},\ldots,\lambda_{k,r_k}\right)$ is the power allocation matrix. While \cite{SwindleBD} computes $\mathbf{\Lambda}_k$ using the classic waterfilling algorithm in order to minimize the power required to achieve rate $R_k$, we can instead apply any of the other schemes discussed in Sec.~\ref{sec:BlindPrec} such as FWF. Due to the subadditivity of the $\rank$ function, reducing the rank of the per-user transmit covariances via FWF effectively reduces the rank of the overall UCT transmit covariance $\mathbf{Q}_{s}$, which in turn mitigates the interference caused to the PR according to Proposition 1.


\section{Primary Outage Probability}\label{sec:PR_OutP}
In this section we characterize the impact of the classic and frugal waterfilling methods on the primary receiver performance assuming independent Rayleigh fading on all channels. Herein, the channels are mutually independent and are each composed of i.i.d. zero-mean circularly symmetric complex Gaussian entries, i.e., $\vect\left(\mathbf{G}_1\right) \sim\mathcal{CN}(\mathbf{0},\mathbf{I})$, and the same distribution holds for $\mathbf{H}_1$,$\mathbf{H}_2$, and $\mathbf{G}_2$.

A first approach would be to directly analyze the PR rate outage probability $I_r=\Pr \left( {{R_p}\left( {{{\mathbf{Q}}_s}} \right) \leq T} \right)$ for a target rate $T$, which is equivalent to
\begin{equation}\label{eq:PRRateOutp}
I_r=\Pr\left({\log _2}\left| {{\mathbf{I}} + {{\mathbf{H}}_1}{{\mathbf{Q}}_p}{\mathbf{H}}_1^H{{{\left( {{{\mathbf{K}}_p} + \sigma _p^2{\mathbf{I}}} \right)}^{ - 1}}}} \right|\leq T \right)
\end{equation}
where ${\mathbf{Q}}_s$ and ${\mathbf{Q}}_p$ are obtained via one of the waterfilling methods on their noise-prewhitened channels and are therefore functions of random matrices $\{\mathbf{G}_1,\mathbf{G}_2\}$ and $\{\mathbf{H}_1,\mathbf{H}_2\}$, respectively. Unfortunately, the computation of (\ref{eq:PRRateOutp}) is prohibitively complex since it is a non-linear function of the eigenvalues of four complex Gaussian random matrices (even if the PT applies uniform power allocation instead), and is an open problem to our best knowledge. Previous studies on the statistical distribution of MIMO capacity under interference usually circumvent this difficulty by assuming the MIMO transmitter and interferer adopt uniform or deterministic power allocation \cite{Lozano02,Moustakas03,Tulino06},
 which reduces the problem to one involving two complex Gaussian random matrices. As such, we are not aware of prior work on the statistics of MIMO capacity under interference where either one or both transmitters employ waterfilling as in our model.

 In light of the above, it is of interest to develop more tractable PR performance measures. One such candidate is the PR interference temperature outage probability (ITOP), which is the probability that ${T_p}\left( {\mathbf{Q}_s} \right)$ [cf. (\ref{eq:IntTemperature})] exceeds a threshold $\eta$:
 \begin{equation}\label{eq:ITOP}
 {I_p}\left( {{{\mathbf{Q}}_s},\eta } \right) = \Pr\left(\Tr\left(\mathbf{H}_{2}\mathbf{Q}_s\mathbf{H}_{2}^H\right) \ge \eta \right).
 \end{equation}
 The ITOP is appealing since the interference temperature metric is widely used in underlay systems, and can be considered to be the MIMO counterpart of efforts to characterize the statistical distribution of aggregate UCT interference in single-antenna networks as in \cite{Clancy07}.
Paradoxically, however, it is seen in Sec.~\ref{sec:sim} that FWF causes the highest ITOP, even though the average primary rate is the highest and the PR rate outage is the lowest when $\mathbf{Q}_s$ is computed using FWF. Therefore, a more accurate surrogate for the PR rate outage $I_r$ is the interference leakage-rate outage probability (ILOP), defined as
 \begin{equation}\label{eq:ILOP}
 {I_l}\left( {{{\mathbf{Q}}_s},\eta } \right) = \Pr\left(\log_2\left|\sigma_p^2\mathbf{I}+\mathbf{H}_{2}\mathbf{Q}_s\mathbf{H}_{2}^H\right| \ge \eta \right),
 \end{equation}
and it is verified in Sec.~\ref{sec:sim} that UCT transmission schemes with the lowest ILOP also minimize $I_r$. This is because the leakage rate has a direct impact on the PR rate: $R_p(\mathbf{Q}_s)$ in \eqref{eq:PURate} can be rewritten as
\begin{eqnarray}\label{eq:Rprewrt}
{R_p}\left( {{{\mathbf{Q}}_s}} \right) &=& {\log _2}\left|\sigma_p^2{\mathbf{I}} + {{\mathbf{H}}_1}{{\mathbf{Q}}_p}{\mathbf{H}}_1^H+\mathbf{K}_p \right| \nonumber\hfill\\
&&\:{-}{\log _2}\left|\sigma_p^2{\mathbf{I}} + \mathbf{K}_p \right|
\end{eqnarray}
where the first term is the sum rate of the virtual PT/UCT multiple access channel (MAC) with optimal successive detection, and the second term is the leakage rate from the UCT. For the worst-case scenario where the PT is decoded first in the virtual MAC, decreasing the leakage rate improves the detection of the PT signal in the first term and simultaneously reduces the second term, thereby decreasing $I_r$.
On the other hand, the link between interference temperature and PR rate is more tenuous.

Assume the UCT transmit covariance matrix $\mathbf{Q}_s$ is of rank $k$, $1 \leq k \leq \min \left( {{N_a},{N_s}} \right)$, where $k$ is determined by the choice of waterfilling scheme to achieve rate $R_b$ over the pre-whitened UCT channel $\mathbf{\tilde G}_1\triangleq {{\left( {{{\mathbf{K}}_s} + \sigma _s^2{\mathbf{I}}} \right)}^{ -1/2}}\mathbf{G}_1$. Assume $\mathbf{\tilde G}_1^H\mathbf{\tilde G}_1$ is of rank $d'$, with non-zero ordered eigenvalues $\left\{ {{\alpha _i}} \right\}_{i = 1}^{d'}$. Waterfilling yields a diagonal $\mathbf{Q}_s$ with entries \cite{Telatar1999}
\begin{equation}\label{eq:Qs_ii}
{\left[ {{{\mathbf{Q}}_s}} \right]_{i,i}} = \left[\mu  - \frac{\sigma_s^2}{{{\alpha _i}}}\right]^+,\;i=1,\ldots,N_a,
\end{equation}
where the waterfilling level $\mu$ is a function of $P_s$, $\sigma_s^2$, and ${\bm{\alpha }} = \left( {{\alpha _1}, \ldots ,{\alpha _{d'}}} \right)$ \cite{Cioffi06,Zanella09}.
Similar arguments hold for the underlay downlink covariance $\mathbf{Q}_s$ designed for sum-rate target $\sum\nolimits_k R_k$ and aggregate channel ${\mathbf{\tilde G}} = \left[ {\begin{array}{*{20}{c}}
  {\left( {\mathbf{Z}}_1^{ - {1 \mathord{\left/
 {\vphantom {1 2}} \right.
 \kern-\nulldelimiterspace} 2}}{{\mathbf{G}}_{1,1}}{{\mathbf{T}}_1}\right)^T}&{\ldots}&{\left( {\mathbf{Z}}_K^{ - {1 \mathord{\left/
 {\vphantom {1 2}} \right.
 \kern-\nulldelimiterspace} 2}}{{\mathbf{G}}_{K,1}}{{\mathbf{T}}_K}\right)^T}
\end{array}} \right]$.

\subsection{Interference Temperature Outage Probability}
Noting that the ITOP and ILOP (\ref{eq:ITOP})-(\ref{eq:ILOP}) are still functions of four complex Gaussian matrices, we first develop bounds on the ITOP as follows. We assume $\mathbf{H}_2$ is full rank such that $\rank(\mathbf{H}_2)=d=\min(N_r,N_a)$, which holds with probability 1 under i.i.d. Rayleigh fading. Define $r=\min\left( {{k},{d}} \right)$ and let $\lambda _i\left( \mathbf{A} \right)$ denote the $i^{th}$ ordered eigenvalue of $\mathbf{A}$ in descending order. Starting with the commutativity of the trace operator,
\begin{eqnarray}
  {I_p} &=& \Pr\left(\Tr\left( {{\mathbf{H}}_2^H{{\mathbf{H}}_2}{{\mathbf{Q}}_s}} \right) \geq \eta\right)\hfill \\
    &\leq& \Pr \left( {\Tr\left( {{\mathbf{H}}_2^H{{\mathbf{H}}_2}{{\mathbf{Q}}_{s|{{\mathbf{Q}}_p} = \left( {{{{P_p}} \mathord{\left/
 {\vphantom {{{P_p}} {{N_p}}}} \right.
 \kern-\nulldelimiterspace} {{N_p}}}} \right){\mathbf{I}}}}} \right) \geq \eta } \right)\label{eq:QpUPA}\\
  &=&\Pr\left(\sum\limits_{i = 1}^r {{\lambda _i}\left( {{\mathbf{H}}_2^H{{\mathbf{H}}_2}{{\mathbf{Q}}_s}} \right)}\geq \eta\right)  \hfill \\
   &\approx&  \Pr \left( {\sum\limits_{i = 1}^r {{\lambda _i}\left( {{\mathbf{H}}_2^H{{\mathbf{H}}_2}{{{\mathbf{\tilde Q}}}_s}} \right)}  \geq \eta } \right)\label{eq:Qsbar}\\
   &\leq& \Pr\left(\sum\limits_{i = 1}^r {{\lambda _i}\left( {{\mathbf{H}}_2^H{{\mathbf{H}}_2}} \right){\lambda _i}\left( {{{\mathbf{\tilde Q}}_s}} \right)}\geq \eta\right)\label{eq:ProdIneq}\\
   &\leq& \Pr\left({\lambda_1}\left( {{{\mathbf{\tilde Q}}_s}}\right)\sum\limits_{i = 1}^r {{\lambda _i}\left( {{\mathbf{H}}_2^H{{\mathbf{H}}_2}}  \right)}\geq \eta\right)\label{eq:Ipbnd}
   \end{eqnarray}
 where in (\ref{eq:QpUPA}) we eliminate dependence on $\mathbf{H}_1$ by assuming the PT adopts uniform power allocation $({{\mathbf{Q}}_p} = \left( {{{{P_p}} \mathord{\left/
 {\vphantom {{{P_p}} {{N_p}}}} \right.
 \kern-\nulldelimiterspace} {{N_p}}}} \right){\mathbf{I}})$ such that ${{\mathbf{K}}_s} = \left( {{{{P_p}} \mathord{\left/
 {\vphantom {{{P_p}} {{N_p}}}} \right.
 \kern-\nulldelimiterspace} {{N_p}}}} \right){{\mathbf{G}}_2}{\mathbf{G}}_2^H$, which is a worst-case interference scenario at the UCR according to Proposition 1 and potentially increases the power expended by the UCT; in (\ref{eq:Qsbar}) ${{{\mathbf{\tilde Q}}}_s}$ is the statistical waterfilling solution where $\mu$ in (\ref{eq:Qs_ii}) is a function of the statistics of $\left\{ {{\alpha _i}} \right\}_{i = 1}^d$ and offers nearly the same performance as instantaneous waterfilling \cite{Cioffi06}-\cite{Zanella09};
%
and the inequality in (\ref{eq:ProdIneq}) follows from the bound on the trace of a product of Hermitian matrices \cite{Zhang_TAC}.

We now define the ordered eigenvalue vectors ${\mathbf{h}} = \left( {{\lambda_1}\left( {{\mathbf{H}}_2^H{{\mathbf{H}}_2}} \right), \ldots ,{\lambda_r}\left( {{\mathbf{H}}_2^H{{\mathbf{H}}_2}} \right)} \right)$ and ${\mathbf{q}} = \left( {{\lambda_1}\left( {\mathbf{\tilde Q}}_s \right), \ldots ,{\lambda_r}\left( {\mathbf{\tilde Q}}_s \right)} \right)$.
Observe that the overall joint density of these random eigenvalues is given by the product of the individual joint densities: ${f_{{\mathbf{h}},{\mathbf{q}}}}\left( {{\mathbf{h}},{\mathbf{q}}} \right) = {f_{\mathbf{h}}}\left( {\mathbf{h}} \right){f_{\mathbf{q}}}\left( {\mathbf{q}} \right)$ due to the independence of the associated channel matrices. Thus, the bound in \eqref{eq:Ipbnd} can be rewritten as
\begin{equation}\label{eq:Exp_Ipbnd}
{I_p} \leq {E_{\mathbf{h}}}\left\{ {1 - {F_{{q_1}}}\left( {{\eta  \mathord{\left/
 {\vphantom {\eta  {\sum\nolimits_{i = 1}^r {{h_i}} }}} \right.
 \kern-\nulldelimiterspace} {\sum\nolimits_{i = 1}^r {{h_i}} }}} \right)} \right\}
 \end{equation}
 where ${F_{{q_1}}}(x)$ is the cumulative distribution function (cdf) of the largest eigenvalue $q_1$.
From \eqref{eq:Qs_ii} we obtain ${F_{{q_1}}}\left( x \right) = {F_{{\alpha _1}}}\left( {{{\sigma _s^2} \mathord{\left/
 {\vphantom {{\sigma _s^2} {\left( {\tilde \mu  - x} \right)}}} \right.
 \kern-\nulldelimiterspace} {\left( {\tilde \mu  - x} \right)}}} \right)$ where $\bar\mu$ is the statistical water level. The cdf of $\alpha_1$, the largest eigenvalue of $\mathbf{\tilde G}_1^H\mathbf{\tilde G}_1$, is given next for the scenario $N_s\geq N_a,N_s\geq N_p$.

 \begin{lemma}\label{lemma:LargestWEigCDF}\cite{Stuber08}
 Given complex Gaussian matrices $\mathbf{X}\in \mathbb{C}^{N_s \times N_a}$,$\mathbf{Y}\in \mathbb{C}^{N_s\times N_p}$, and $(N_p\times N_p)$ diagonal matrix $\mathbf{P}=\diag(\rho,\ldots,\rho)$, the cdf of the largest eigenvalue $\alpha_{max}$ of the quadratic form ${{\mathbf{X}}^H}\left( {{\mathbf{YP}}{{\mathbf{Y}}^H} + {\sigma ^2}{\mathbf{I}}} \right){\mathbf{X}}$ when $N_s\geq N_a,N_s\geq N_p$ is
 \begin{equation}\label{eq:cdf_alpha1}
 {F_{{\alpha _{max }}}}\left( x \right) = {K_1}\left| {{{{\mathbf{\tilde \Delta }}}_1}\left( {x} \right)} \right|
 \end{equation}
 where $ {{{{\mathbf{\tilde \Delta }}}_1}\left( {x} \right)}  = {\left[ {\begin{array}{*{20}{c}}
  {{\mathbf{\tilde Y}}{{\left( {x} \right)}^T}}&{{{{\mathbf{Z}}}^T}}
\end{array}} \right]^T}$,
\[{\left[ {{{\mathbf{\tilde Y}}}\left( {{x }} \right)} \right]_{i,j}} = \left\{ {\begin{array}{*{20}{c}}
  \begin{gathered}
  \Gamma \left( i \right){I_{{N_a} - i}}\left( \rho  \right) - \Gamma \left( i \right){e^{ - x}} \hfill \\
   \times \sum\limits_{k = 0}^{i - 1} {\frac{{{x^k}}}{{k!}}{I_{{N_a} - i}}\left( {\frac{\rho }{{1 + \rho x}}} \right),\; i = 1, \ldots ,{N_a},}  \hfill \\
\end{gathered}  \\
  {{{\left( { - 1} \right)}^{{N_s} - j}}{I_{{N_a} + {N_s} - i}}\left( \rho  \right),\; i = {N_a} + 1, \ldots ,{N_s},}
\end{array}} \right.\]
 $I_a\left( {b} \right) = \sum\nolimits_{k = 0}^a {\left( {\begin{array}{*{20}{c}}
  a \\
  k
\end{array}} \right){b^{j + k}}\Gamma \left( {j + k} \right)}$, and the normalization constant $K_1$ \cite[eq. (25)]{Stuber08} and the entries ${\left[ {{\mathbf{Z}}} \right]_{i,j}}$ \cite[eq. (27)]{Stuber08} are functions of the array dimensions independent of $x$.
 \end{lemma}

The cdf ${F_{{\alpha _{1 }}}}\left( x \right)$ for other antenna array dimensions is of a similar form and can be found in \cite{Stuber08}.
Now, in order to compute the expectation over $\mathbf{h}$ in \eqref{eq:Exp_Ipbnd}, we exploit the Gaussian distribution of $\mathbf{H}_2$ based on the following lemma.

\begin{lemma}\label{lemma:jointWisheigpdf}\cite{Telatar1999,Chiani03}
If $\mathbf{X}$ is a $(N_r \times N_a)$ matrix with i.i.d. zero-mean unit-variance complex Gaussian elements, then $\mathbf{W}={\mathbf{XX}}^H$ follows a central complex Wishart distribution if $N_r \leq N_a$, otherwise $\mathbf{W}={\mathbf{X}^H\mathbf{X}}$ is Wishart-distributed if $N_r > N_a$.
Given $c\triangleq\max(N_r,N_a)$ and $m\triangleq\min(N_r,N_a),$ the joint density of all $m$ ordered eigenvalues of ${{\mathbf{W}}}$ is
\begin{equation}\label{eq:jointWisheigpdf_uncorr}
{f_{\bm{\Lambda }}}\left( {{\lambda _{1}}, \ldots ,{\lambda _{{m}}}} \right) = K_w{\left| {{{\mathbf{V}}_1}\left( {\bm{\lambda }} \right)} \right|^2}\prod\limits_{i = 1}^{{m}} {\frac{e^{ - {\lambda _{i}}}}{\left({\lambda _{i}}\right)^{{m}-{c}}}},
\end{equation}
where ${{{\mathbf{V}}_1}\left( {\bm{\lambda }} \right)}$ is a Vandermonde matrix with entries ${\left[ {{{\mathbf{V}}_1}\left( {\mathbf{\lambda }} \right)} \right]_{i,j}} = \lambda _j^{i - 1}$,
and $K_w$ is a normalization constant independent of $\bm{\lambda}$ \cite[eq. 7]{Chiani03}.
\end{lemma}

For the case where $\rank(\mathbf{\tilde Q}_s)=k$ is greater than $\rank(\mathbf{H}_2)=d$, i.e., $r=d$, the term ${\sum\nolimits_{i = 1}^r {{h_i}} }$ in \eqref{eq:Exp_Ipbnd} involves all $d$ ordered eigenvalues $\left\{ {{h_1}, \ldots ,{h_d}} \right\}$ and the associated joint density function $f_{\mathbf{h}}\left( {\mathbf{h}} \right)$ is given in Lemma~\ref{lemma:jointWisheigpdf}. Thus \eqref{eq:Exp_Ipbnd} yields
\begin{equation}\label{eq:DblIntegrald}
{I_p} \leq 1 - K\int { \ldots \int\limits_\mathfrak{D} {\left| {{{{\mathbf{\tilde \Delta }}}_1}\left( {{{\bar h}}} \right)} \right|} } \left| {{{\mathbf{V}}_1}\left( {\mathbf{h}} \right)} \right|\prod\limits_{i = 1}^d {\xi \left( {{h_i}} \right)} d{\mathbf{h}}
\end{equation}
 where $K={K_1}{K_w}$, ${{{\bar h}}}={{{\sigma _s^2} \mathord{\left/
 {\vphantom {{\sigma _s^2} {\left( {\tilde \mu  - {\eta  \mathord{\left/
 {\vphantom {\eta  {\sum\nolimits_{i = 1}^d {{h_i}} }}} \right.
 \kern-\nulldelimiterspace} {\sum\nolimits_{i = 1}^d {{h_i}} }}} \right)}}} \right.
 \kern-\nulldelimiterspace} {\left( {\tilde \mu  - {\eta  \mathord{\left/
 {\vphantom {\eta  {\sum\nolimits_{i = 1}^d {{h_i}} }}} \right.
 \kern-\nulldelimiterspace} {\sum\nolimits_{i = 1}^d {{h_i}} }}} \right)}}}$, $\xi \left( {{h_i}} \right)={\frac{e^{ - {h_{i}}}}{\left({h_{i}}\right)^{{m}-{c}}}}$, and the integration region is $\mathfrak{D} = \left\{ {\infty \geq {h_1} \geq {h_2} \geq  \ldots  \geq {h_d} \geq 0} \right\}$. This multidimensional integral has a closed-form solution obtained from the following generalized Cauchy-Binet identity.

 \begin{lemma}\label{lemma:Cauchy-Binet}\cite[Lemma 2]{Chiani06}
For $\mathbf{x}=\left\{ {{x_1}, \ldots ,{x_M}} \right\}$, arbitrary integrable functions $c_i(\cdot)$, $u_i(\cdot)$, and ${\varphi \left( \cdot \right)}$, $N \times N$ matrix ${{\mathbf{\Phi }}\left( {\mathbf{x}} \right)}$ and $M \times M$ matrix ${{\mathbf{\Psi }}\left( {\mathbf{x}} \right)}$ ($M\leq N$), where $\Phi \left( {\mathbf{x}} \right) = {\left[ {\begin{array}{*{20}{c}}
  {{{\mathbf{C}}_1}{{\left( \mathbf{x} \right)}^T}}&{{{\mathbf{C}}_2}^T}
\end{array}} \right]^T}$, with entries ${{{\left[ {{{\mathbf{C}}_1}\left( {\mathbf{x}} \right)} \right]}_{i,j}} = {c_i}\left( {{x_j}} \right)}$ for $i=1,\ldots,N-M;j=1,\ldots,N$, ${{{\left[ {{{\mathbf{C}}_2}} \right]}_{i,j}} = {c_{2i,j}}}$ (constant scalars) for $i=N-M+1,\ldots,N;j=1,\ldots,N$, and ${{{\left[ {{{\mathbf{\Psi}}}\left( {\mathbf{x}} \right)} \right]}_{i,j}} = {u_i}\left( {{x_j}} \right)}$, the following integral identity over domain $\mathfrak{D} = \left\{ {b \geq {x_1} \geq {x_2} \geq  \ldots  \geq {x_N} \geq a} \right\}$ holds:
\begin{equation}
  \int { \cdots \int\limits_\mathfrak{D} {\left| {{\mathbf{\Phi }}\left( {\mathbf{x}} \right)} \right| \cdot \left| {{\mathbf{\Psi }}\left( {\mathbf{x}} \right)} \right|} } \prod\limits_{k = 1}^N {\varphi \left( {{x_k}} \right)} d{\mathbf{x}}
   = M! \det \mathbf{B}
\end{equation}
where
\begin{equation}
{\left[ {\mathbf{B}} \right]_{i,j}} = \left\{ {\begin{array}{*{20}{c}}
  {\int_a^b {\varphi \left( x \right){c_j}\left( x \right){u_i}\left( x \right)dx} },\text{}{i=1,\ldots,N-M;\forall j} \\
  {{c_{2i,j}}},\quad{i=N-M+1,\ldots,N;\forall j.}
\end{array}} \right.
\end{equation}
\end{lemma}

A compact solution to \eqref{eq:DblIntegrald} is then obtained by setting $\mathbf{\Phi }={{{\mathbf{\tilde \Delta }}}_1}\left( {{{\bar h}}} \right)$ with $\rho=P_p/N_p$, ${{\mathbf{\Psi }}\left( {\mathbf{x}} \right)}={{{\mathbf{V}}_1}\left( {\mathbf{h }} \right)}$, and $\varphi \left( {{x}} \right)=\xi \left( {{h}} \right)$ in Lemma~\ref{lemma:Cauchy-Binet}:
\begin{equation}\label{eq:Ip1}
{I_p} \leq 1 - d!K\left| {\mathbf{B}_1} \right|;
\end{equation}
\begin{equation}\label{eq:B_1-ij}
{\left[ {\mathbf{B}_1} \right]_{i,j}} = \left\{ {\begin{array}{*{20}{c}}
  {\int_0^\infty {\xi \left( x \right)x_j^{i-1} {\left[ {{{\mathbf{\tilde Y}}}\left( {{x }} \right)} \right]_{i,j}} dx} },\text{}{i=1,\ldots,N-M;\forall j} \\
  {\left[ {{\mathbf{Z}}} \right]_{i,j}},\quad{i=N-M+1,\ldots,N;\forall j.}
\end{array}} \right.
\end{equation}
and requires a computationally-inexpensive one-dimensional numerical integration of the product of elementary functions in \eqref{eq:B_1-ij}.

On the other hand, when $r=k<d$, only the $k$ largest eigenvalues are included in the term ${\sum\nolimits_{i = 1}^r {{h_i}} }$ in \eqref{eq:Exp_Ipbnd}, which necessitates invoking the corresponding joint density function given below in Lemma~\ref{lemma:jointWisheigsubsetpdf}.
\begin{lemma}\label{lemma:jointWisheigsubsetpdf}\cite{Win09}
The joint density function of the ordered subset of the $s$ largest eigenvalues of Wishart matrix $\mathbf{W}$ having $m$ non-zero eigenvalues in total is
\begin{equation}\label{eq:Lemma4}
\begin{gathered}
{f_{{\lambda _1}, \ldots ,{\bm{\lambda} _s}}}\left( {\mathbf{\lambda }} \right) = K_w\sum\limits_{{\mathbf{n}},{m},s} \sum\limits_{{\mathbf{m}},{m},s} sgn\left( {{\mathbf{n}},{\mathbf{m}}} \right)\left| {\mathbf{D}\left( {{\lambda _s}} \right)} \right|\\
\quad\qquad\times \prod\limits_l^s e^{-1}\lambda _l^{{N_r} - {N_a} + {n_l} - {m_l} + 2}
\end{gathered}
\end{equation}
where $\mathbf{n}=\mathbf{m}=\{1,\ldots,s\}$, each summation is a $N_r$-fold nested sum over permutations ${r_{i,{\mathbf{n}}}},{r_{j,{\mathbf{m}}}}$ of the index sets as defined in \cite[eq. (16)]{Win09} with sign determined by $sgn\left( {{\mathbf{n}},{\mathbf{m}}} \right)\in \left\{ { \pm 1} \right\}$ \cite[eq. (17)]{Win09},
\begin{equation}
{\left[ {\mathbf{D}\left( {{\lambda _s}} \right)} \right]_{i,j}} = \gamma \left( {{N_a} - {N_r} + {r_{i,{\mathbf{n}}}} + {r_{j,{\mathbf{m}}}},{\lambda_s}} \right),
\end{equation}
and $\gamma(a,b)$ is the incomplete gamma function \cite{Stuber08}.
\end{lemma}

Substituting $\varphi \left( {{x_k}} \right)=x_k^{{N_r} - {N_a} + {n_l} - {m_l} + 2}$ and associated expressions into \eqref{eq:Exp_Ipbnd} and invoking Lemma~\ref{lemma:Cauchy-Binet} provides
\begin{equation}\label{eq:Ip2}
{I_p} \leq 1 - \frac{k!K}{e^k}\sum\limits_{{\mathbf{n}},{d},k} \sum\limits_{{\mathbf{m}},{d},k} sgn\left( {{\mathbf{n}},{\mathbf{m}}} \right)\left| {\mathbf{B}_2} \right|
\end{equation}
where
\begin{equation}\label{eq:B_2-ij}
{\left[ {\mathbf{B}_2} \right]_{i,j}} = \left\{ {\begin{array}{*{20}{c}}
  {\int_0^\infty {\varphi \left( {{x_k}} \right){\left[ {\mathbf{D}\left( {{x}} \right)} \right]_{i,j}} {\left[ {{{\mathbf{\tilde Y}}}\left( {{x }} \right)} \right]_{i,j}}dx} },\text{}{i=1,\ldots,d-k,} \\
  {\left[ {{\mathbf{Z}}} \right]_{i,j}},\quad{i=d-k+1,\ldots,N;\forall j.}
\end{array}} \right.
\end{equation}
Since $\gamma(a,b)$ is a standard function in MATLAB, the numerical integration of its product with two elementary functions in \eqref{eq:B_2-ij} is straightforward.

\subsection{Interference Leakage Outage Probability}
Turning our attention to the ILOP, starting with the definition of $I_l$ we have
\begin{eqnarray}
  {I_l}\left( {{{\mathbf{Q}}_s},{\mathbf{\eta }}} \right) &=& \Pr \left( {\sum\nolimits_{i = 1}^r {{{\log }_2}\left( {\sigma_p^2 + {\lambda _i}\left( {{\mathbf{Q}}_s}{\mathbf{H}}_2^H{{{\mathbf{H}}_2}} \right)} \right) \geq \eta } } \right) \nonumber\hfill \\
   &\approx& \Pr \left( {{{\log }_2}\prod\nolimits_{i = 1}^r {{\lambda _i}\left( {{\mathbf{Q}}_s}{\mathbf{H}}_2^H{{{\mathbf{H}}_2}} \right) \geq \eta } } \right) \label{eq:InterfLimit}\hfill \\
   &\leq& \Pr \left( {\prod\nolimits_{i = 1}^r {{\lambda _i}\left( {{{{\mathbf{\tilde Q}}}_s}} \right){\lambda _i}\left( {{{\mathbf{H}}_2}{\mathbf{H}}_2^H} \right) \geq {2^\eta }} } \right) \label{eq:prod} \hfill \\
   &\leq& \Pr \left( {{{\left( {{q_1}} \right)}^r}\prod\nolimits_{i = 1}^r {{h_i} \geq {2^\eta }} } \right)\\
   &=& {E_{\mathbf{h}}}\left\{ {1 - {F_{{q_1}}}\left( {{2^\eta  \mathord{\left/
 {\vphantom {2^\eta  {\prod\nolimits_{i = 1}^r {{h_i}} }}} \right.
 \kern-\nulldelimiterspace} {\prod\nolimits_{i = 1}^r {{h_i}} }}} \right)^{1/r}} \right\}\label{eq:Exp_Ilbnd}
\end{eqnarray}
where in (\ref{eq:InterfLimit}) we consider the interference-limited scenario which is of interest, and (\ref{eq:prod}) is due to \cite{Zhang_TAC}.

The computation of \eqref{eq:Exp_Ilbnd} closely parallels that of \eqref{eq:Exp_Ipbnd}, by again separating the cases $r=d<k$ and $r=k<d$, followed by invoking Lemmas~\ref{lemma:LargestWEigCDF}--\ref{lemma:Cauchy-Binet} for the former and Lemmas~\ref{lemma:LargestWEigCDF},\ref{lemma:jointWisheigsubsetpdf} and \ref{lemma:Cauchy-Binet} for the latter case, respectively. Therefore, the resulting closed-form bounds for the ILOP are of the form in \eqref{eq:Ip1} and \eqref{eq:Ip2}, with $\bar h=\left( {{2^\eta  \mathord{\left/
 {\vphantom {2^\eta  {\prod\nolimits_{i = 1}^r {{h_i}} }}} \right.
 \kern-\nulldelimiterspace} {\prod\nolimits_{i = 1}^r {{h_i}} }}} \right)^{1/r}$ and all other terms being unchanged.


\section{Numerical Results}\label{sec:sim}
In this section, we present some numerical examples to demonstrate the performance of the proposed rank-minimization UCT transmit covariance designs in MIMO cognitive radio networks. We consider MIMO cognitive radio networks with one primary user and one or more underlay receivers. In all simulations, the channel matrices and background noise samples were assumed to be composed of independent, zero-mean Gaussian random variables with unit variance. In situations where the desired rate for UCT cannot be achieved with the given $P_s$, rather than indicate an outage, we simply assign all power to transmit signals. The performance is evaluated by averaging over $1000$ independent channel realizations.

\subsection{Single Underlay Receiver Scenario}
\begin{figure}[htbp]
\centering
\includegraphics[width=3.5in]{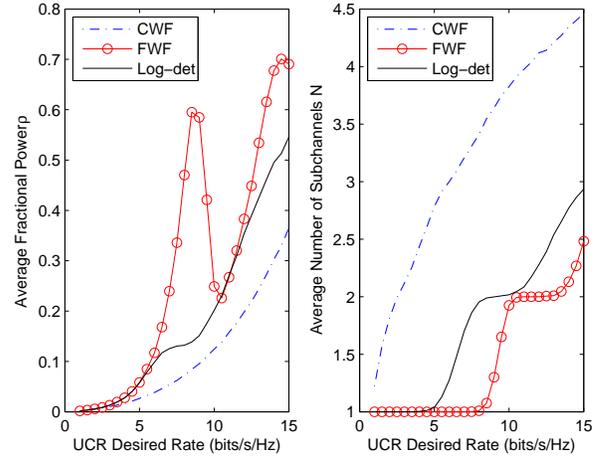}
\caption{Power and dimension allocation versus UCR desired rate.}\label{fig:fig1}
\end{figure}
We first consider the single UCR scenario, where each node is equipped with $6$ antennas, and $P_s=100$, $P_p=10$. Fig.~\ref{fig:fig1} illustrates the average fractional power $\rho$ and the average number of subchannels $N$ required to achieve the UCR desired rates by \emph{CWF}, \emph{FWF} and \emph{log-det heuristic} algorithms. It is shown that the trace and rank of the UCT transmit covariance matrix $\mathbf{Q}_s$ are two competing objectives, and any scheme which requires more power occupies fewer spatial dimensions. Among the three methods, CWF demands the largest spatial footprint, while the FWF scheme offers the smallest feasible number of transmit dimension. We should point out that the log-det heuristic algorithm for matrix rank minimization does not always provide the smallest transmit dimension, compared to FWF. This is because the log-det algorithm is an approximate heuristic and can only give a local minimum.

\begin{figure}[htbp]
\centering
\includegraphics[width=3.5in]{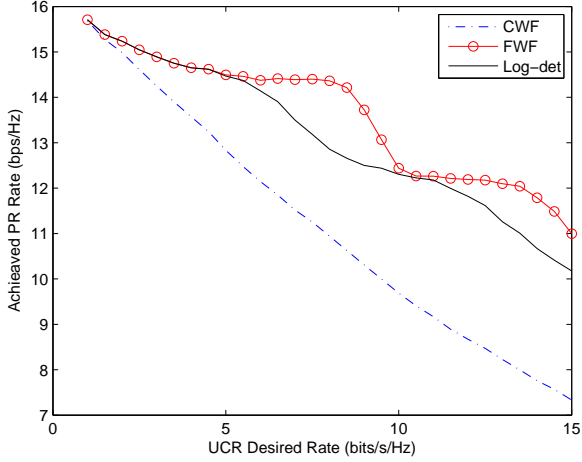}
\caption{Achieved PU rate versus UCT desired rate.}\label{fig:fig2}
\end{figure}
The achieved average primary user data rates for all the methods is depicted in Fig.~\ref{fig:fig2}. As expected, the lower-rank UCT transmit convariance will cause lesser degradation on average to the PU communication link, thus resulting a higher PR rate in accordance with Proposition 1. Compared to CWF, either of the proposed modified waterfilling algorithms or the log-det heuristic lead to more desirable PR rates, with the advantage of FWF being more pronounced as $R_b$ increases.

\begin{figure}[htbp]
\centering
\includegraphics[width=3.5in]{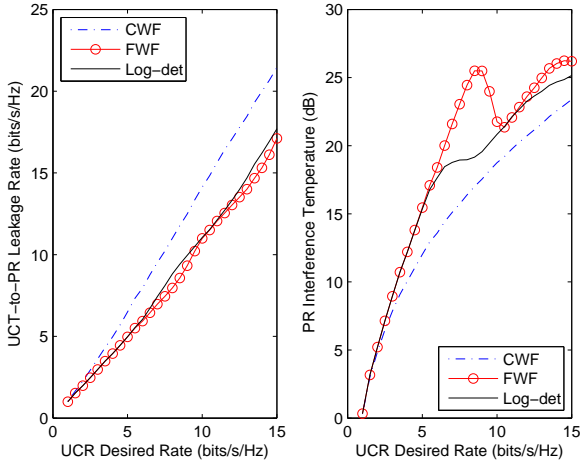}
\caption{Two metrics of PR Interference versus UCR desired rate.}\label{fig:fig3}
\end{figure}
To obtain greater insight, Fig.~\ref{fig:fig3} compares two metrics of interference at PR using different algorithms, where one is the newly-defined UCT-PR leakage rate, the other one is the commonly-used interference temperature. We notice that an interesting phenomenon: the two metrics gives the opposite trend. It is worth to point out that the commonly-used interference temperature metric does not accurately capture the interference impact caused by the UCT on the primary mutual information, while the interference leakage rate remedies this defect.

\begin{figure}[htbp]
\centering
\includegraphics[width=3.3in]{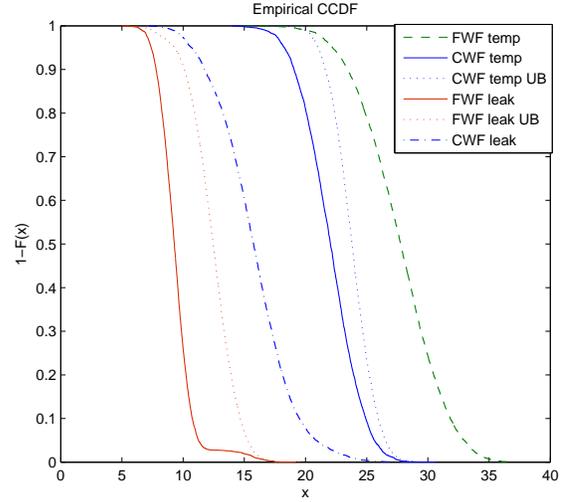}
\caption{Empirical ccdf of interference temperature and leakage rate under CWF and FWF.}\label{fig:LeakTempccdf}
\end{figure}
\begin{figure}[htbp]
\centering
\includegraphics[width=3.5in]{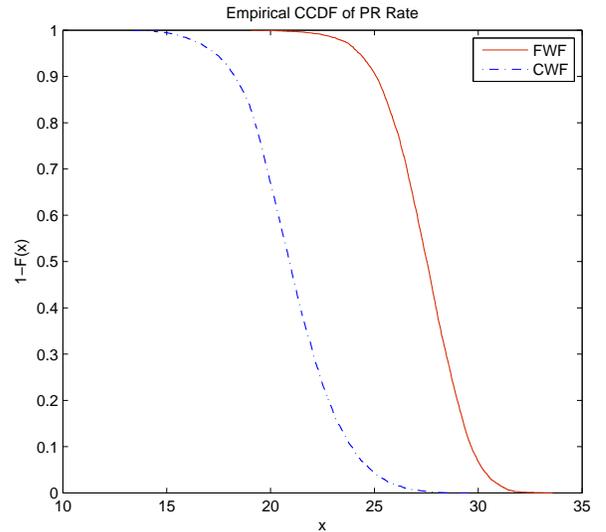}
\caption{Empirical ccdf of PR rate under CWF and FWF.}\label{fig:PRrateccdf}
\end{figure}

For the statistical characterization of the proposed schemes, we exhibit the empirical complementary cdfs and select analytical upper bounds from Sec.~\ref{sec:PR_OutP} of the interference temperature and leakage rate metrics in Fig.~\ref{fig:LeakTempccdf}, for 6 antennas at all users and $P_s=200,R_b=8,P_p=40$. An immediate observation is the conflicting trends of the leakage and temperature metrics: FWF causes a much greater interference temperature outage and much smaller leakage rate outage compared to CWF, and the superiority of one versus the other is not apparent. To resolve this dilemma, the corresponding empirical PR rate ccdfs are shown in Fig.~\ref{fig:PRrateccdf}, and it is clear that employing FWF leads to a very significant reduction in PR rate outage probability as compared to CWF. Furthermore, the interference temperature outage is again seen to be misleading regarding the true impact on the PR rate outage probability. Thus, FWF outperforms CWF in terms of both average PR rate and PR rate outage probability.

\subsection{MIMO Underlay Downlink Scenario}
\begin{figure}[htbp]
\centering
\includegraphics[width=3.5in]{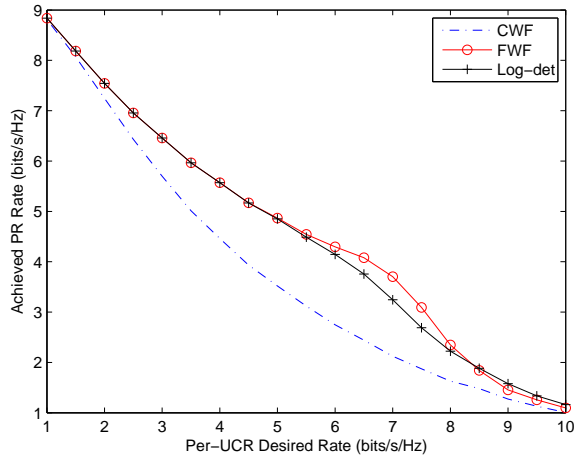}
\caption{Achieved PR rate versus per-UCR desired rate in MIMO underlay downlink, $P_s=20dB$, $P_p=10dB$.}\label{fig:fig5}
\end{figure}
Next, we evaluate the performance of the proposed algorithms with the modified BD strategy of Sec.~\ref{sec:CRdownlink} for a MIMO underlay downlink system, where there are $K=3$ UCRs, and $N_a=12$, $N_p=N_r=N_s=4$. Without loss of generality it is assumed that the desired rate targets for all UCRs are the same, i.e. $R_1=R_2=R_3$. Fig.~\ref{fig:fig5} illustrates the achieved PU rate versus per UCT desired rate, when $P_s=100$ or $20dB$ and $P_p=10$ or $10dB$. The benefit of minimizing the transmit covariance rank is seen to hold even for the multi-user downlink scenario.

\begin{figure}[htbp]
\centering
\includegraphics[width=3.5in]{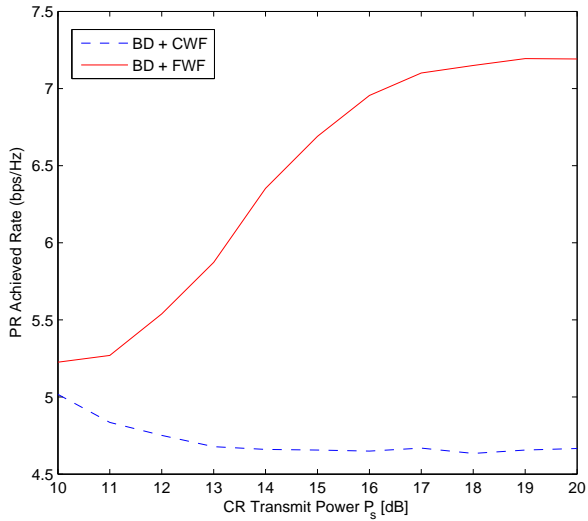}
\caption{Achieved PR rate versus UCT transmit power in MIMO downlink system with identical target rates $R_1=R_2=R_3=5$bits/s/Hz.}\label{fig:fig6}
\end{figure}
\begin{figure}[htbp]
\centering
\includegraphics[width=3.5in]{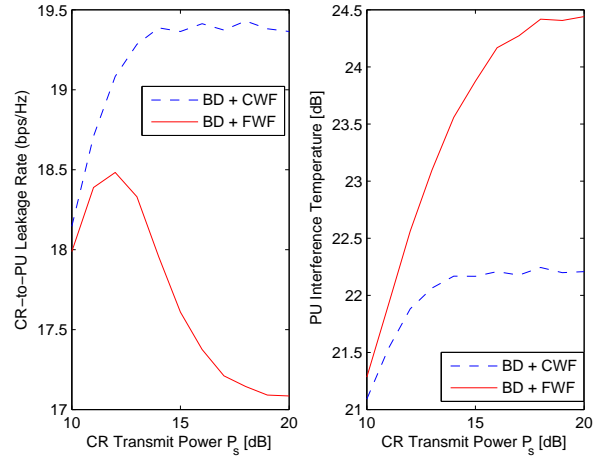}
\caption{Two metrics of PR interference versus UCT transmit power in MIMO underlay downlink with identical target rates $R_1=R_2=R_3=5$.}\label{fig:fig7}
\end{figure}

It is also of interest to see how the achieved PR rates under the various designs vary with the UCT transmit power, when the desired rate for each UCT is fixed. The simulation settings are the same as above, except that we fix $R_1=R_2=R_3=5$ and $P_s \in [10dB,20dB]$. The results are shown in Fig.~\ref{fig:fig6}, with the corresponding leakage rate and interference temperature metrics in Fig.~\ref{fig:fig7}. Once again, FWF offers the optimal average PR rate and PR rate outage probability in the downlink scenario.

\section{Conclusion}\label{sec:concl}
This paper has proposed a rank minimization precoding
strategy for underlay MIMO CR systems with completely
unknown primary CSI, assuming a minimum information rate
must be guaranteed on the CR main channel. We presented a
simple waterfilling approach can be used to find the minimum
rank transmit covariance that achieves the desired CR rate
with minimum power. We also presented two alternatives to
FWF that are based on convex approximations to the minrank
criterion, one that leads to conventional waterfilling for
our CR problem, and another based on a log-det heuristic.
The CWF approach turns out to be a poor approximation to
the min-rank objective, while the log-det approach provides
performance similar to FWF, although FWF consistently leads
to the highest throughput for the primary link. We also
observed that reducing the inteference temperature metric is
surprisingly not consistent with improving the PR throughput;
in particular, FWF has the highest interference temperature of
the algorithms studied, but also leads to the highest PR rate.
As an alternative, we proposed an interference leakage metric
that is a better indicator of the impact of the CR on the primary
link.


%
%
%
%
%
\end{document}